\documentclass[twoside]{dis08}
\usepackage[latin1]{inputenc}
\usepackage[dvips]{graphicx,epsfig,color}
\usepackage{wrapfig,rotating}
\usepackage{amssymb,amsmath,array}

\begin{document}
\title{Leptoquarks and Contact Interactions at LeHC}

\author{Aleksander Filip \.Zarnecki 
%
%
\vspace{.3cm}\\
%
 Institute of Experimental Physics, University of Warsaw \\
   Ho\.za 69, 00-681 Warszawa, Poland
%
}

\maketitle

\begin{abstract}
The sensitivity of LeHC to different models of ``new physics''
has been studied, both for the resonance production and 
in the contact interaction approximation.
Expected limits are compared for different running scenarios.
Direct leptoquark production can be studied for masses up to about 2~TeV.
For contact interaction models scales up to about 70~TeV 
can be explored.
Significant improvement of existing limits is also expected for models
with large extra dimensions.
Effective Plank mass scales up to about 5.4 TeV can be probed.
LeHC will be sensitive to the quark substructure of the order of
$10^{-19}$~m.
\end{abstract}


\section{Introduction}

Search for "new physics" has always been one of the most 
important subjects in the field of particle physics. 
The first electron-proton collider HERA proved to be well suited
both for precise tests of the Standard Model and for 
constraining its possible extensions.
Large Hadron electron Collider (LeHC) is the proposal for the
next-generation electron-proton machine.
It would bring 7~TeV proton beam of the LHC into collisions 
with high-energy electrons (or positrons).
Both linac and ring options for the electron machine are considered.

Presented analysis was developed in 2000/2001 as a contribution
to TESLA TDR and the THERA Book~\cite{therabook}.
Same approach (with only minor modifications) has been used since 2005
to demonstrate physics capabilities of the $ep$ upgrade option of the LHC.
Results presented here were obtained assuming four 
different running scenarios for LeHC:
\begin{itemize}
\item lepton beam energy of 70 GeV, 
integrated luminosities of $2\times 10$ or $2\times 100$ fb$^{-1}$,
\item lepton beam energy of 140 GeV, 
integrated luminosities of $2\times 1$ or $2\times 10$ fb$^{-1}$.
\end{itemize}
For details of the analysis method and the considered models
the reader is referred to the previous study~\cite{appb}.


\section{Analysis method}

In this contribution production of leptoquark states, as classified  
by Buchm\"uller, R\"uckl and Wyler \cite{pl:b191:442}, are considered.
For leptoquark masses, $M_{LQ}$, smaller than the available $ep$ 
center-of-mass energy direct production of single leptoquarks 
can be studied.
If leptoquark Yukawa coupling, $\lambda_{LQ}$, is of the order 
of the electroweak coupling or smaller, the resonance width is small
compared to the expected precision of mass reconstruction.
In such a case the narrow-width approximation (NWA) can be used to
describe the cross-section for single leptoquark production 
in electron-proton (for leptoquarks with fermion number $F=2$)
or positron-proton (for $F=0$) scattering:
\begin{eqnarray}
\sigma^{ep\rightarrow LQ \; X}(M_{LQ},\lambda_{LQ}) & = & 
(J+1) \cdot \frac{\pi \lambda_{LQ}^{2}}{4 M_{LQ}^{2}} 
        \cdot x_{LQ} q(x_{LQ},M_{LQ}^{2})
\nonumber
\end{eqnarray}
where $q(x,Q^{2})$ is the quark momentum distribution
in the proton and $x_{LQ} =\frac{M_{LQ}^{2}}{s}$.
Resonance production should manifest itself by a narrow peak in
the electron-jet invariant mass distribution for high $Q^2$ NC DIS events.
To suppress SM background contribution additional cut on the $y$ variable, 
related to the $eq$ scattering angle in the center-of-mass frame, is imposed.
Assuming that the measured distribution comes from the SM processes only, 
expected limits on $\lambda_{LQ}$ as a function of $M_{LQ}$ can be calculated
for different running scenarios.

In the limit of heavy leptoquark masses ($M_{LQ} \gg \sqrt{s}$)
the effect of leptoquark production or exchange
is equivalent to a vector type $eeqq$ contact interaction (CI).
The influence on the NC $ep$ DIS cross-section can be described 
by introducing additional terms $\eta^{eq}_{ij}$
in the tree level $eq \rightarrow eq$ scattering amplitudes:
\begin{eqnarray}
\eta^{eq}_{ij} & = & 
a^{eq}_{ij} \cdot \left( \frac{\lambda_{LQ}}{M_{LQ}} \right)^{2} 
\nonumber
\end{eqnarray}
where the subscripts $i$ and $j$ label the chiralities of the initial 
lepton and quark respectively ($i,j=L,R$) and 
the coefficients $a^{eq}_{ij}$ depend on the leptoquark type.
In the CI method limits on the leptoquark mass to the Yukawa coupling ratio
are extracted from the measured $Q^2$ distribution of NC DIS events.
For leptoquark masses comparable with the $ep$ center-of-mass energy
modified CI approach can be used. 
Dependence of the  $\eta^{eq}_{ij}$ terms 
on the process kinematics is included, separately for
$u$-channel leptoquark exchange process and the $s$-channel production
contribution.

\begin{figure}[t]
\centerline{
\includegraphics[width=0.51\columnwidth]{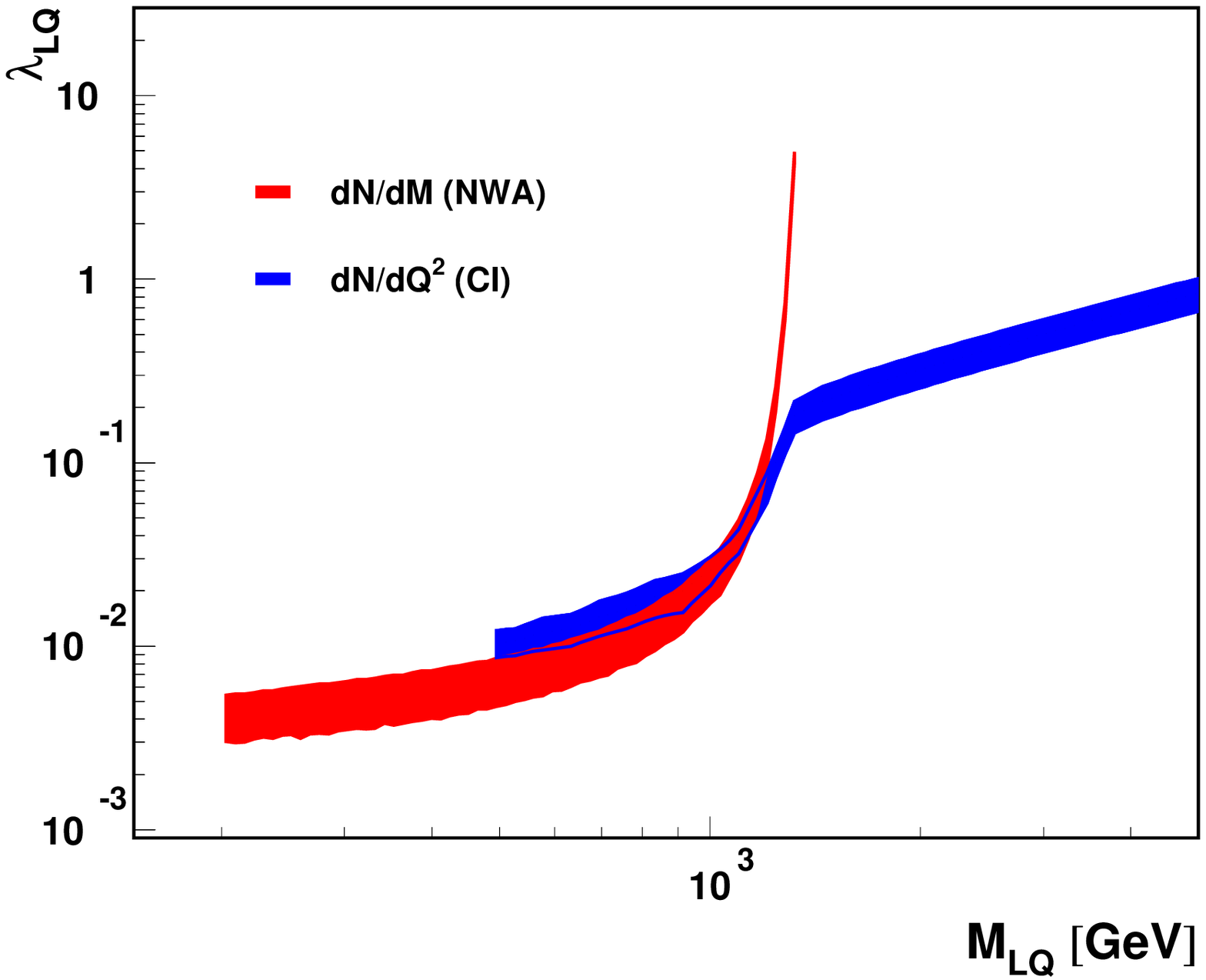}
\includegraphics[width=0.47\columnwidth]{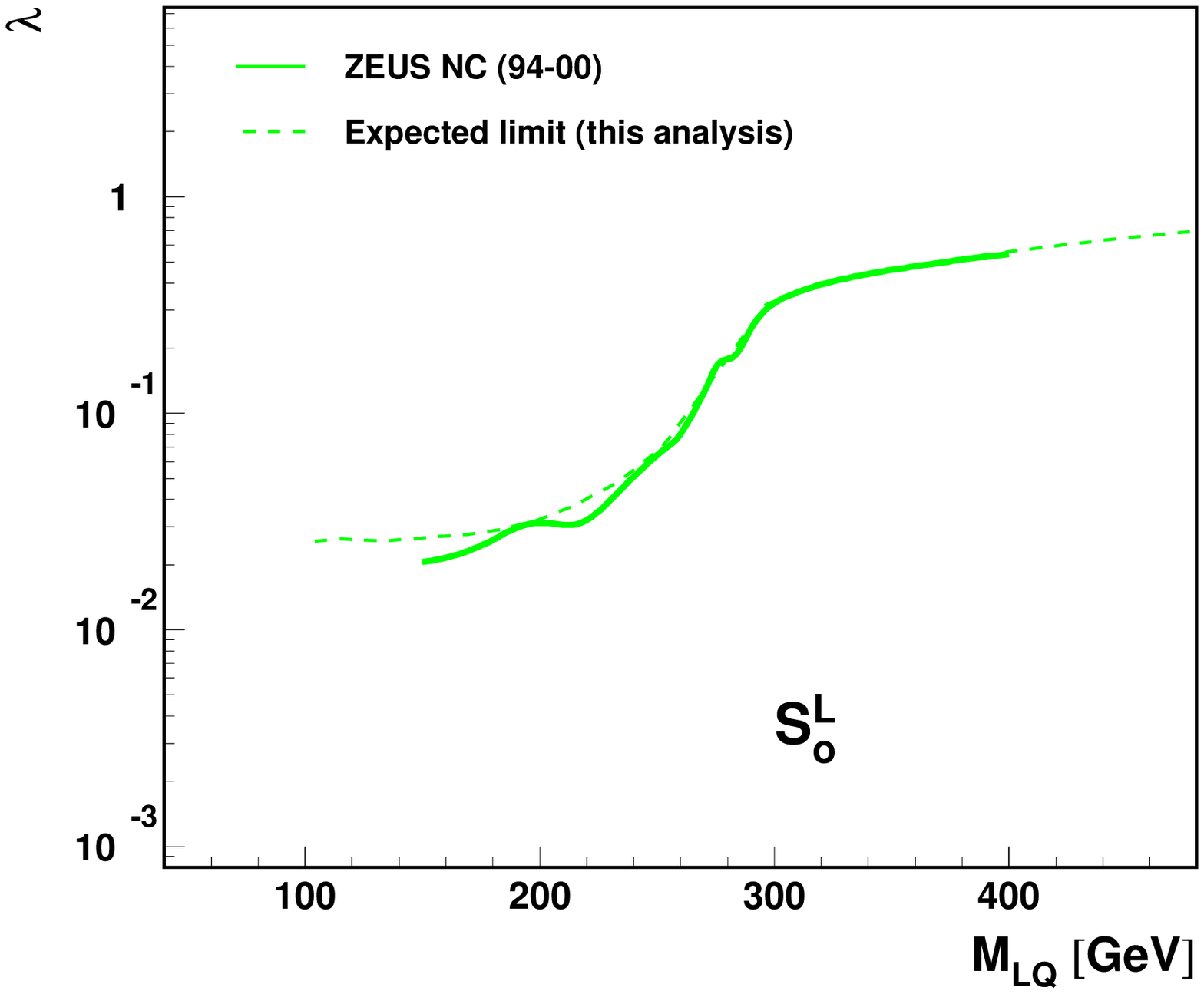}
}
\caption{
Left: comparison of limits for $S^L_0$ leptoquark 
obtained with NWA and in modified CI approach, as expected for LeHC 
running with 70~GeV lepton beam energy and luminosity 
of $2\times 10$ fb$^{-1}$.
The width of the limit curves shows the level of expected statistical 
fluctuations, as obtained from simulation of multiple MC experiments.
Right: results of the ZEUS collaboration~\cite{zeuslq} compared with 
the limits expected for given luminosity calculated 
with presented approach. 
}\label{fig:method}
\end{figure}
Modified CI approach can be used also for $M_{LQ} < \sqrt{s}$.
However, for low leptoquark masses better constraints are obtained from 
NWA approach.
Shown in Figure~\ref{fig:method}a is the comparison of limits on
Yukawa coupling of $S^L_0$ leptoquark, as expected at LeHC from 
invariant mass distribution measurement (NWA) and form $Q^2$ distribution 
measurement (CI method) of NC DIS events.
Both kinds of limits are always calculated for leptoquark masses 
$M_{LQ} < \sqrt{s}$ and the stronger one is taken.
Presented approach has been verified by comparison with published ZEUS
results~\cite{zeuslq}, see Figure ~\ref{fig:method}b.


\section{Results}

\begin{figure}[t]
\centerline{
\includegraphics[width=0.49\columnwidth]{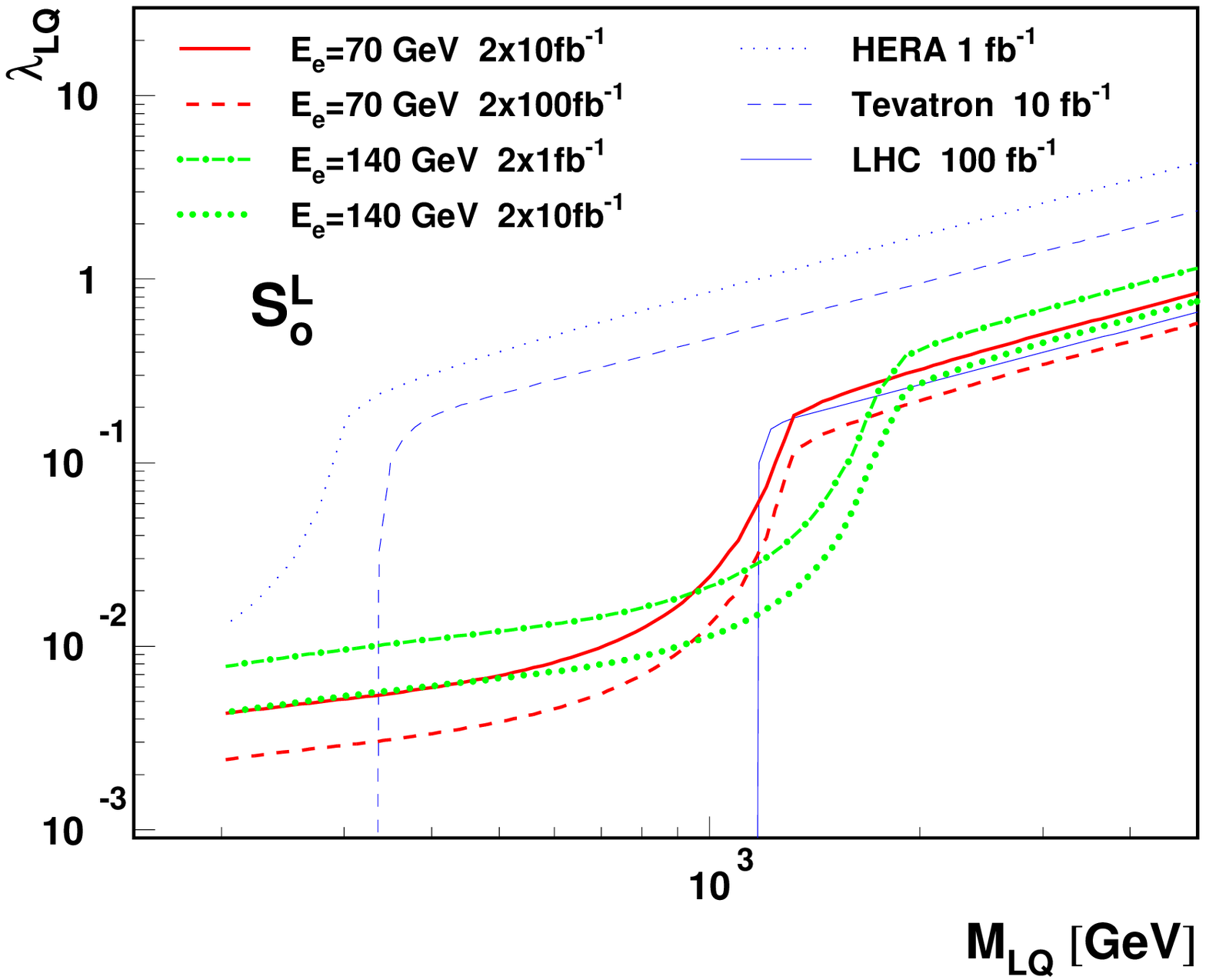}
\includegraphics[width=0.49\columnwidth]{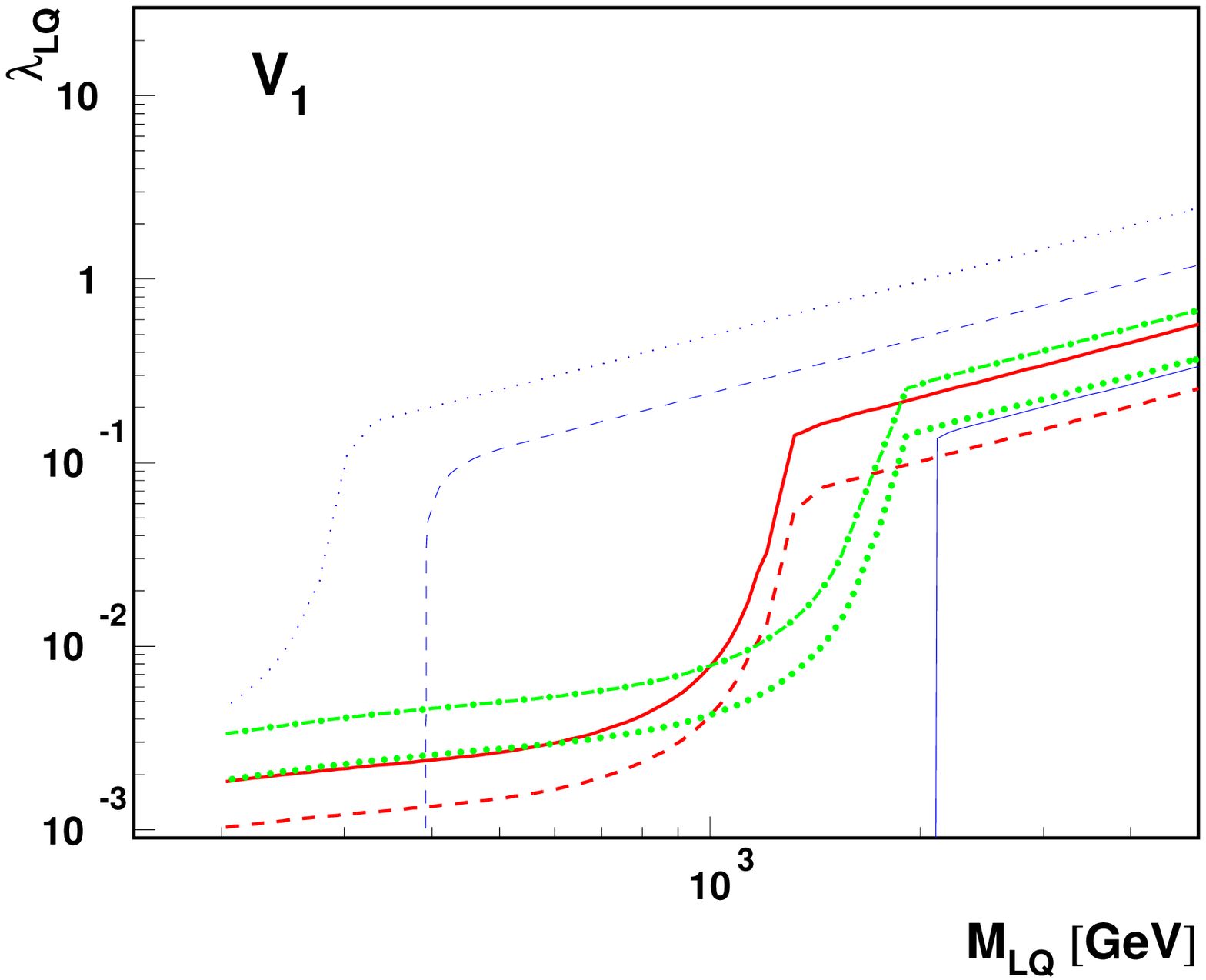}
}
\caption{Expected 95\% CL exclusion limits in  $(\lambda_{LQ},M_{LQ})$, 
for considered LeHC running scenarios, for $S^L_0$ and $V_1$ leptoquark models.
Expected final limits from HERA, the Tevatron 
and the LHC are included for comparison.
}\label{fig:lq}
\end{figure}

Shown in Figure~\ref{fig:lq} are expected 95\% CL exclusion limits 
in  $(\lambda_{LQ},M_{LQ})$, for selected leptoquark models and 
four considered LeHC running scenarios.
Expected final limits from HERA, the Tevatron and the LHC are also indicated.
Discovery reach of LeHC is not larger than that of LHC.
However, if any leptoquark-like states are discovered at the LHC, 
with masses below 1~TeV, they could be precisely studied at LeHC.

\begin{figure}[htb]
\centerline{
\includegraphics[width=0.42\columnwidth]{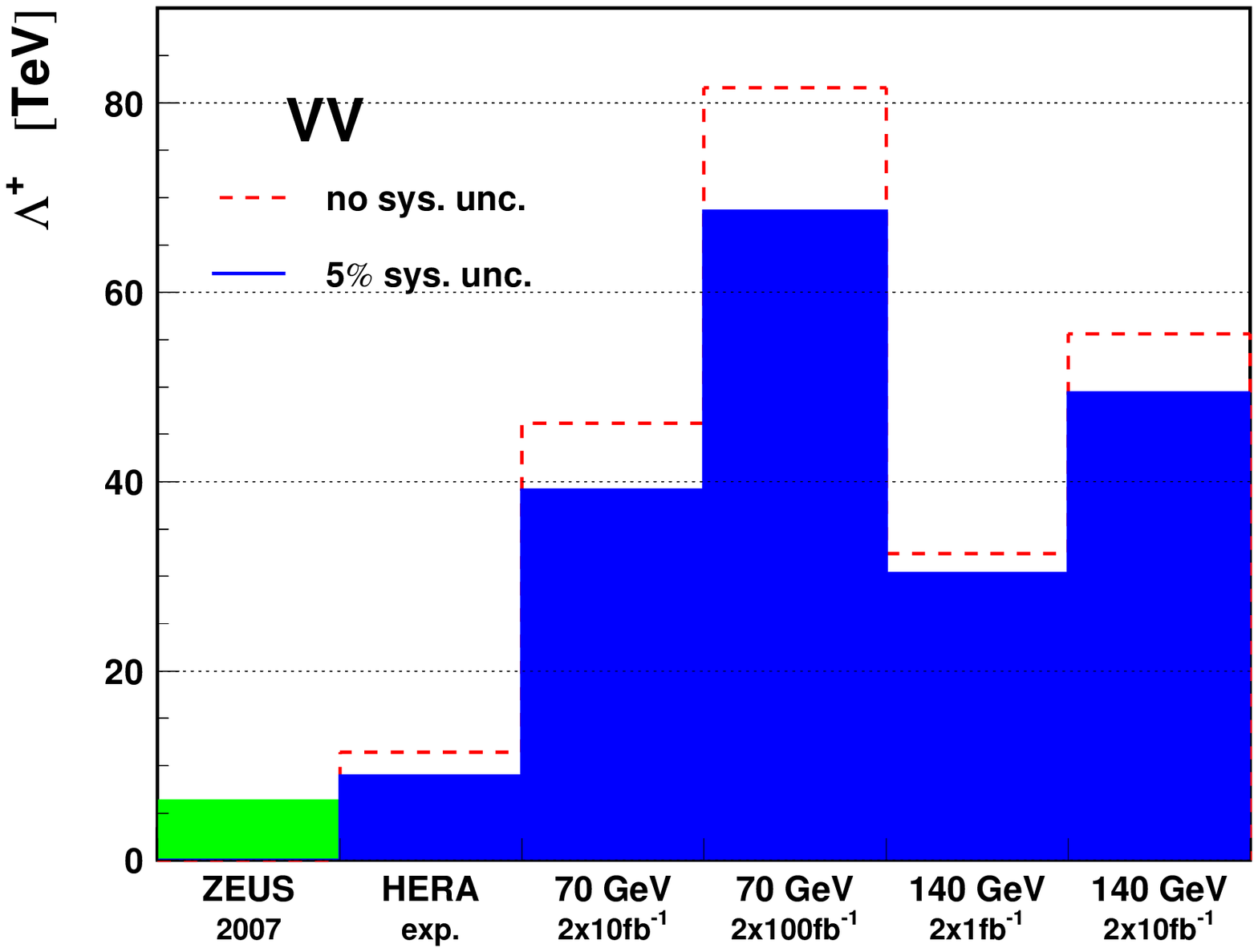}
\includegraphics[width=0.42\columnwidth]{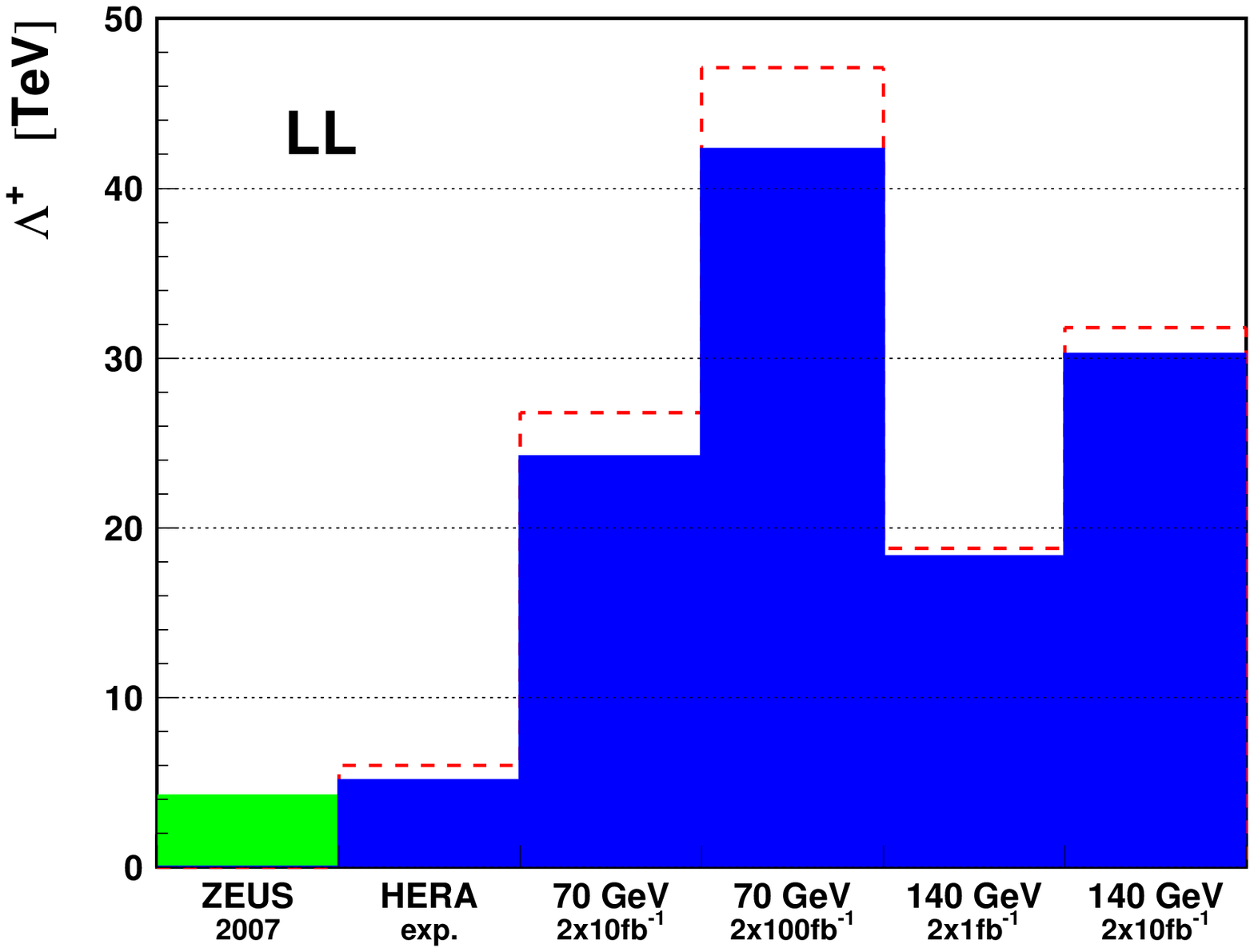}
}
\caption{95\% CL exclusion limits on the mass scales
$\Lambda^{+}$ expected from the measurement of high-$Q^{2}$ NC DIS 
cross-sections at LeHC, for the general contact interaction 
models VV and LL.}\label{fig:ci}
\end{figure}

CI approach can also be used to constrain other extensions of the SM. 
for which low energy effects coming from "new physics" 
at much higher energy scales can be approximated by four-fermion
contact interactions.
Shown in Figure~\ref{fig:ci} are 95\% CL exclusion limits on the mass scales
$\Lambda^{+}$ 
expected from the measurement of high-$Q^{2}$ NC DIS cross-sections 
at LeHC for the general contact interaction models VV and LL.
Current ZEUS limits~\cite{zeusci} 
and expected final HERA limits are included for comparison.
LeHC will be sensitive to CI mass scales of the order of 50~TeV.
However, uncertainties of SM expectations and other systematic effects 
have to be reduced to few \% level.

\begin{figure}[htb]
\centerline{
\includegraphics[width=0.42\columnwidth]{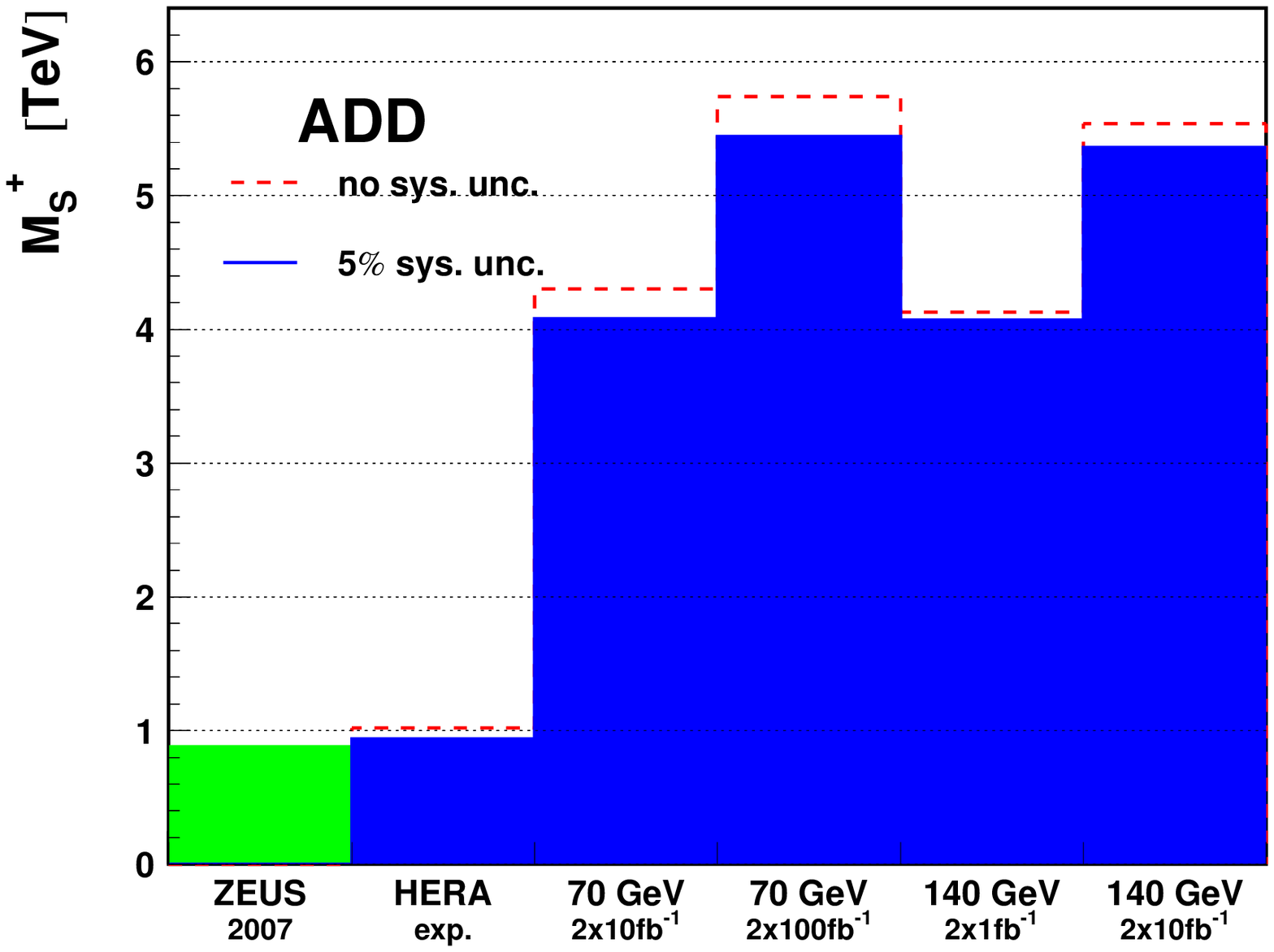}
\includegraphics[width=0.42\columnwidth]{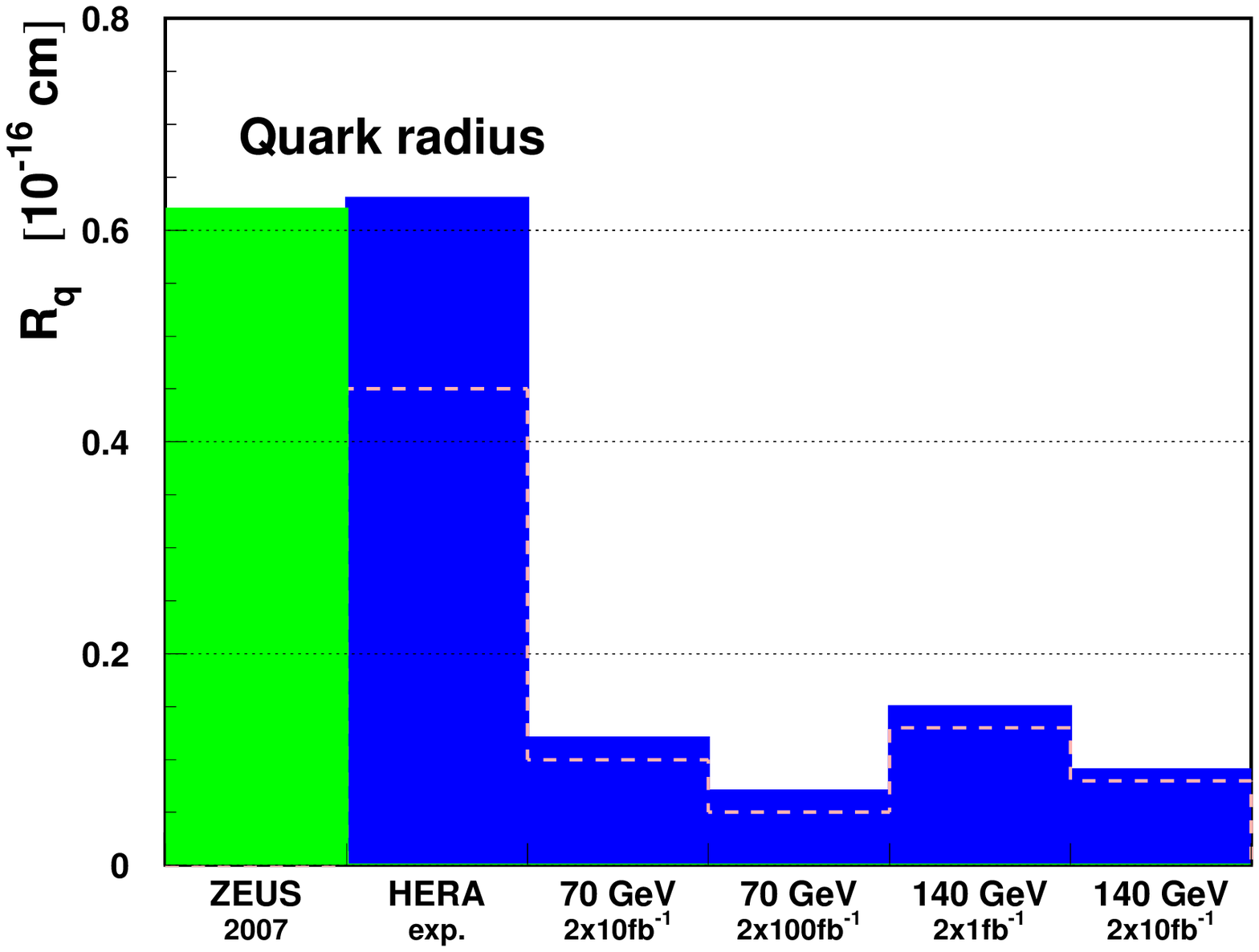}
}
\caption{95\% CL exclusion limits on the effective Planck 
mass scale in models with large extra dimensions (left) and on
the effective quark charge radius (right) expected
from the measurement of high-$Q^{2}$ NC DIS cross-sections 
at the LeHC.}\label{fig:add}
\end{figure}

Corresponding limits on the effective Planck mass scale, $M_S$, in model with 
large extra dimensions~\cite{add} and on the effective quark charge 
radius, $R_q$, are presented in Figure~\ref{fig:add}.
LeHC will be sensitive to the mass scales of extra dimensions of the order
of 4--5~TeV.
For the effective quark-charge radius, in the classical form-factor
approximation, constraints below $10^{-19}$~m can be obtained.



\begin{footnotesize}

\end{footnotesize}


\end{document}